# Experimenting with Convolutional Neural Network architectures for the automatic characterization of Solitary Pulmonary Nodules' malignancy rating

Ioannis D. Apostolopoulos[1]

Department of Medical Physics, School of Medicine, University of Patras, Greece, ece7216@upnet.gr

**Abstract**

Lung cancer is the most common cause of cancer-related death worldwide. Early and automatic diagnosis of Solitary Pulmonary Nodules (SPN) in Computed Tomography (CT) chest scans can provide early treatment as well as doctor liberation from time-consuming procedures. Deep learning has been proved as a popular and influential method in many medical imaging diagnosis areas. In this study, we consider the problem of diagnostic classification between benign and malignant lung nodules in CT images, derived from a PET/CT scanner. More specifically, we intend to develop experimental Convolutional Neural Network (CNN) architectures and conduct experiments, by tuning their parameters, to investigate their behavior, and define the optimal setup for the accurate classification. For the experiments, we utilize PET/CT images obtained from the Laboratory of Nuclear Medicine of University of Patras, and the public database called Lung Image Database Consortium Image Collection (LIDC-IDRI). Furthermore, we apply simple data augmentation to generate new instances and inspect the performance of the developed networks. Classification accuracy of 91% and 93% on PET/CT and on a selection of nodule images from the LIDC-IDRI datasets is achieved accordingly. The results demonstrate that CNNs are a trustworthy method for nodule classification. Also, the experiment confirms that data augmentation enhances the robustness of the CNNs.

**Keywords**

Convolutional Neural Networks, Deep Learning, Solitary Pulmonary Nodule Classification, SPN two-type classification, Data Augmentation

# 1. Introduction

Lung cancer is the most common cause of cancer death in men, and the second leading cause of cancer death in women, globally [1]. Recent researches estimate that approximately 1.6 million lung cancer deaths occurred in 2012 worldwide, accounting for about 19 % of all cancer deaths [2].

Low-dose lung Computer Tomography (CT) screening provides an effective way for early diagnosis. Early diagnosis can significantly reduce the lung cancer mortality rate. Compared to other imaging techniques, Computerized Tomography can visualize small or low-contrast nodules. This enhances our ability to diagnose cancer-related Solitary Pulmonary Nodules (SPN) at an early stage. Early detection of Solitary Pulomonary Nodules from lung CT scans can lead to diagnosis and early remedy of lung cancer.

Commonly, clinicians observe, analyze, and interpret the CT scans according to the results of nodule morphology and clinical conditions [3]. However, this procedure has some issues. Firstly, the number of cases a doctor has to examine on a daily basis is very large. Secondly, the interpretation of each lesion by radiologists is a complicated process. Finally, an inefficient analysis may cause patients to miss the best time for treatment. The diagnosis is subjective, and clinicians' professional levels differ. Different clinicians may give diverse interpretations according to their level of experience [4]. In addition, due to the human's physical factors such as the limitation of the visual system, fatigue, and distraction, clinicians may not make the best use of the CT image data. Mining knowledge from medical images is not an easy procedure, while it is considered a challenging task. Most of the time, the experts may be unaware of hidden information that may come with a CT scan.

Machine learning and Deep Learning have become established disciplines in applying artificial intelligence to mine, analyze, and recognize patterns from data. Reclaiming the advances of those fields to the benefit of clinical decision making and computer-aided systems is increasingly becoming mandatory, as new large-scale datasets emerge [6].

Deep learning alludes to a broad class of Machine Learning methods and structures, with the sign of utilizing a massive number of hidden layers [7]. The layers process nonlinear information. Each layer involves a transformation of the data into a higher and more abstract level. The deeper we go into the network, the more complex information is learned. More top layers enhance parts of the information that are significant for segregation and smother irrelevant attributes. Usually, deep learning refers to deeper networks than classic machine learning ones.

Convolutional neural networks (CNN) have proven to be powerful tools for a broad range of computer vision tasks [8]. Since 2010, the annual ImageNet Large Scale Visual Recognition Challenge (ILSVRC) has brought dramatic progress in image processing. This competition is based on the ImageNet database, which contains over 14 million images, belonging to 1000 classes, and annotated by the human hand. Many impressive CNNs have been proposed over the last years [9-13], aiming to accurately classify those images.

Before Deep Learning became popular, manual feature engineering followed by classifiers was the general pipeline for nodule classification [14]. After the large-scale LIDC-IDRI [15] dataset became publicly available, deep learning-based methods have become the dominant framework for nodule classification research.

Wang et al. define some drawbacks and concerns regarding the publicly available datasets for the classification task of lung nodules [16]. Their aim is to highlight the need for new, clearly labeled, and larger-scale datasets in order for the research to develop further. The new researchers are looking towards overpassing that obstacle by designing algorithms and techniques that do not rely on large scale data [16].

In comparison with LIDC-IDRI, the PET/CT dataset utilized for this study contains a fewer number of nodules. However, the characterization of the nodules is more trustworthy, since it was based not only on empirical examination by radiologists but on biopsies, nodule examination and FGD consumption computed from PET/CT scanner.

In this work, we investigate the performance of three proposed network architectures on the classification task in the abovementioned datasets.

Firstly, we design a sequential network and perform experiments, by tuning the parameters of the model, making modifications to the network's architecture, and tuning hyperparameters, to achieve optimal results. Then, we construct a dual-path network and investigate its capabilities to generalize. We observe that overfitting is reduced, with an accuracy cost on the initial dataset. Finally, we construct a three-path network for the same task, to inspect other parameters and their contributions. For the evaluation of the CNNs, we follow a specific strategy. Firstly, we test the developed CNNs on folds of the dataset utilized for the training process. Secondly, we test the developed CNNs on the alternative CT dataset, which was completely hidden during the training process. In this way, we investigate CNN's ability to generalize. For the first evaluation, 10-fold cross-validation is performed. During the experiments, we slightly augment the datasets and inspect the data augmentation effect on the performance.

Based on the results, we come up with some observations and concussions, which need future research and investigation. Besides, an encouraging accuracy is obtained on both datasets. Specifically, the accuracy of the PET/CT dataset reached 91%, while the accuracy of the LIDC-IDRI dataset reached 93%.

## 2. Materials and Methods

### 2.1. Convolutional Neural Networks

Convolutional Neural Networks (CNNs) are used extensively in computer vision applications. Recently, they were employed for medical image recognition and image processing tasks, with remarkable results [16]. They were introduced by Le Cun [17], while they are of growing reputation, especially during the last five years [18].

CNN is a deep neural network, which utilizes convolution and pooling layers, in order to extract useful information from the input data, feeding a final fully connected layer [19-20].

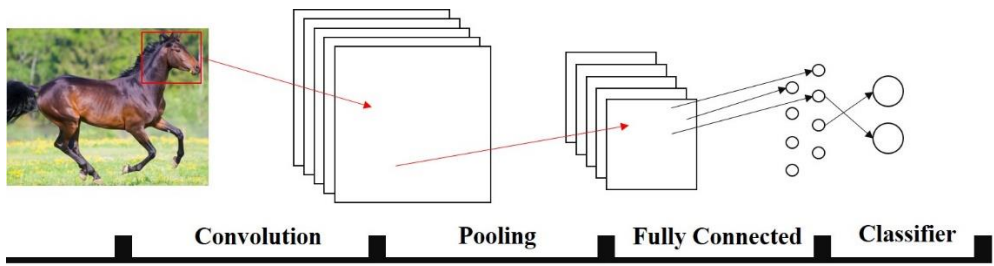

**Figure 1.** Typical Architecture of a CNN. The Convolution Layers operate as feature extractors, while the pooling layers are utilized for dimensionality reduction. The extracted features are classified by a Fully Connected layer (Neural Network).

The architecture shown in Figure 1 allows CNNs to extract increasingly summary representations from the lower layer to the higher layer. Figure 2 [17] illustrates the convolutional and pooling operations.

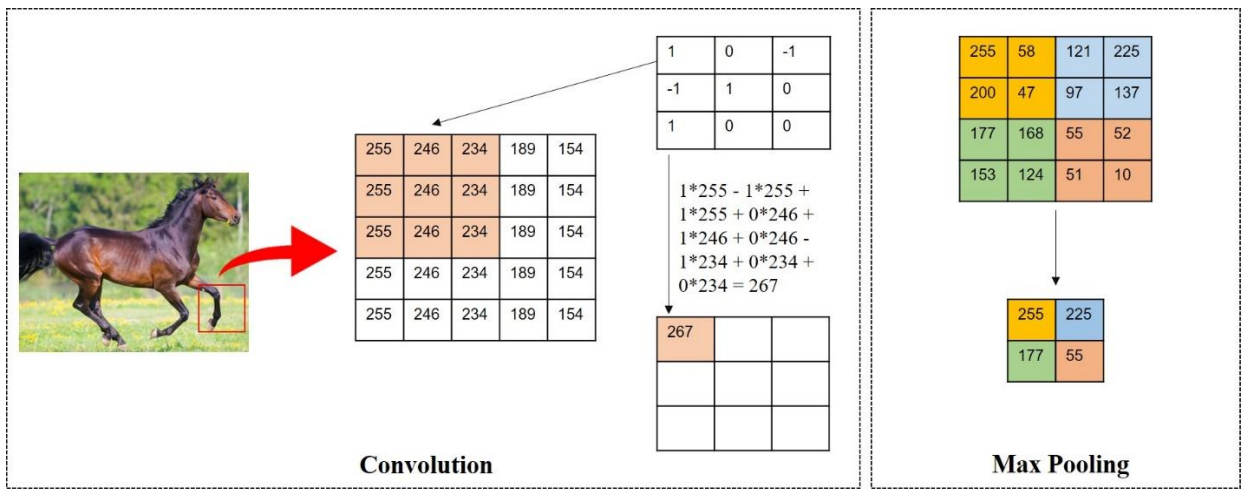

**Figure 2.** Illustration of convolution and max-pooling.

CNNs consist of several parameters, and they are described by complexity. We briefly mention the main characteristics of CNNs, focusing on certain components that we use in our study. We refer to a specific bibliography for an extensive review of each component.

### 2.1.2. Convolution

A convolution operation is performed as a filter is sliding throughout the image. A filter is a table of weights. The size of the table, i.e., the size of the filter, is usually small, concerning the input image size. Common filter sizes, which are called kernels, or receptive fields, are 1x1, 3x3, 5x5, and 7x7. An output pixel produced at every position is a weighted sum of the input pixels (the pixels that the filter has passed from). The weights of the filter are constant for the duration of the scan, thereby extracting information with the same procedure at each iteration. Therefore, convolutional layers can seize the shift-invariance of visible patterns and depict robust features [18]. Convolution filters usually come in batches and are followed by other layers.

More specifically, each neuron of a feature map is connected to a region of neighboring neurons in the previous layer. Such a neighborhood is referred to as the neuron's receptive field in the previous layer. The new feature map can be obtained by convolving the input with a learned kernel and then applying an element-wise activation function on the convolved results. The activation function is briefly explained in the next section. To generate each feature map, the kernel is shared by all spatial locations of the input. The complete feature maps are obtained by utilizing several different kernels. Recently, some improved convolutional formulas are proposed, which have been applied to state-of-the-art networks, with interesting results on specific tasks. Those include dilated convolution [21], tiled Convolution [22], transposed Convolution [23], Network-in-Network [24], and Inception module [9].

### 2.1.3. Activation Function

The activation function is a non-linear transformation, traditionally used in neural networks. For convolutional layers, the activation function is carried out after the convolution process. A proper activation function significantly improves the performance of a CNN for a certain task. One of the most famous and oftenly utilized activation functions is the Rectified Linear Unit (ReLU) [25]. It has been shown that deep networks can be trained efficiently using ReLU even without pre-training [26]. Clevert et al. [27] introduced the Exponential Linear Unit (ELU), which enables faster learning of deep neural networks, and leads to higher classification accuracies in specific tasks and architectures. ELU is a function tending to converge the cost to zero in a faster way, and produce more accurate results. Different from other activation functions, ELU has an extra alpha constant, which should be a positive number. ELU is very similar to RELU except for negative inputs.

Other activation functions include Leaky ReLU [28], Parametric ReLU [29], and Randomized ReLU [30], Maxout [31], and Probout [32]. We note that each activation function enhances or decreases the performance of a network, with respect to the combination of other parameters. Therefore, no activation function fits every situation.

### 2.1.4. Pooling.

The task of the pooling layers is to reduce the dimensions, and consequently, the parameters of the network. The pooling operation fuzzes the properties of the specific place, given that a few spatial features are not significant, thereby reducing the area of interest. It is used widely during the first levels of the network, in order to achieve dramatic dimensionality reduction. One common practice is the distribution of pooling layers among the network, usually after convolutional layers.

The pooling operation computes a specific norm over small areas. This operation aggregates small pitches of pixels and downsamples the photo capabilities from the previous layer, which dramatically reduces the computational cost, especially when the network is deep. The most common pooling operation is max-pooling, which outputs the most pixel value of the area. In most cases, the pooling layer procedure includes max pooling or mean pooling. As far as the mean pooling is concerned, it calculates the average neighborhood of the feature points. Many Pooling techniques have also been suggested, including Lp Pooling [33], Mixed Pooling [34], Stochastic Pooling [35], Spectral pooling [36], Spatial Pyramid Pooling [37], and Multiscale Pooling [38].

### 2.1.5. Loss function.

A loss function is used to evaluate the difference between the output of a CNN, and the class it originally belongs to (i.e., the loss). The minimization of the value of loss function is the main purpose of the training process, as it is used as a marker depicting how adequate the network has learned the training set. This is usually achieved with stochastic gradient descent [39], an optimization technique that calculates the gradient of the loss, with respect to the weight of every part in the network, and then updates the weights according to the computed gradient. Among many types of loss functions, the cross-entropy loss [16], along with Softmax output activations, is the most important loss function for classification tasks [16]. Cross – entropy loss calculates the cross-entropy between the ground truth distribution and the predicted distribution of CNNs. Another commonly used loss function is Hinge loss [40], which measures the difference between the score of the correct class and the score of the predicted class. Other loss functions include Softmax Loss [41], Constrative Loss [42], Triplet loss [43], and Kullback – Leibler Divergence [44].

### 2.1.6. Regularization.

Overfitting is an unneglectable problem in deep CNNs, which can be effectively reduced by regularization. In short, the term overfitting refers to the situation where the network has learned the training set analytically, in a way that it is not able to generalize. Global characteristics are overwhelmed by local and insignificant characteristics that may explain the training set, but cannot describe features that exist in the test sets.

The prevention of overfitting is related to every parameter of the network. For example, overfitting may come from an inadequate architecture, small filter sizes that learn only local characteristics, irrational utilization of MaxPooling layers, etc. The regularization procedure we explain in this section refers to independent techniques used as enhancements for the prevention of overfitting. We explain Dropout and lp-norm techniques as they are the most common methods for this task.

Regularizations referred to lp-norm, modify the objective function by adding terms that apply penalties against the model's over-complexity [16]. There are two kinds of lp-norm regularization, namely l1 norm and l2 norm, sharing the same mathematical approach [45]. A more principled alternative of $l2$-norm regularization is Tikhonov regularization [46], which rewards invariance to noise in the inputs. The mathematic formulas that describe lp-norm regularizers can be found in [16].

Another method for regularization is Dropout, which was introduced by Hinton et al. [45], and has been proven to be very effective in overfitting reduction. Usually, the most significant number of parameters is found in the fully connected layers, on the top of the CNN. Therefore, the most common practice is the application of Dropout layers between fully-connected layers [45].

Dropout can prevent the network from becoming too dependent on any (or any small combination) of neurons and can force the network to be accurate even in the absence of certain information. Several methods have been proposed to improve Dropout. Wang et al. [47] proposed a fast Dropout method which can perform fast Dropout training by sampling from or integrating a Gaussian approximation. Ba et al. [48] proposed an adaptive Dropout method, where the Dropout probability for each variable is computed using a binary belief network that shares parameters with the deep network. In [49], the authors find out that applying standard Dropout before a 1x1 convolutional layer does not prevent overfitting. Therefore, they propose a new Dropout method called Spatial Dropout, which extends the Dropout value across the entire feature map. This new Dropout method works well, especially when the training data size is small.

### 2.1.7. Optimization.

In this sub-section, we briefly discuss data augmentation, weight initialization, stochastic gradient descent, and batch normalization.

<u>Data Augmentation</u>

Deep CNNs are highly dependent on the availability of large-scale training data. An elegant solution to circumvent the shortage of data, compared to the number of parameters involved in CNNs, is data augmentation [26].

Data augmentation involves transforming the available data into new data without altering their nature. Popular augmentation methods include simple geometric transformations, such as sampling [26], mirroring [50], rotating [51], shifting [52], and photometric transformations [53]. Advanced data augmentation procedures have lately been proposed with promising results [16]. The most successful methods include data augmentation via Generative Adversarial Networks (GAN) [54], and data augmentation via Synthetic Minority Oversampling Technique (SMOTE) [55]. The latter method produces new images, which are unrelated to the original ones.

GANs are a particular type of neural network model, where two networks work together simultaneously on two different tasks. The task of the first network, which is also called the generator, is to generate new images that are visually related to the originals, by adding predefined noise. The other network, known as the discriminator, accepts a real or generated sample coming from the generator and assigns a probability that the specific sample is a real or generated one.

SMOTE intends to artificially generate new instances of the class represented by fewer samples, called minority class, using the nearest neighbors of those cases. It is usually combined with manual or automatic under-sampling of the class with the most instances.

Weight initialization

Deep CNNs have a massive number of parameters, which makes the training difficult. Assigning specific values for the weights of the network is an essential prerequisite to achieve faster convergence in training, and to avoid vanishing gradient [56]. The weights parameters initialization should be defined carefully to break the symmetry among hidden units of the same layer. There several methods for random initialization of the weight values, including the proposal by Krizhevsky et al. [26], by Glorot et al. [57], and by Saxe et al. [58].

Stochastic Gradient Descent

The backpropagation algorithm is the standard training method that uses gradient descent to update the parameters. Many gradient descent optimization algorithms have been proposed [26].

Batch Normalization

As the data flows through different layers of a deep network, its distribution of input to internal layers is changed, thereby reducing the learning capacity and the accuracy of the network. Commonly, the first step of data preprocessing involves data normalization. However, this is not enough, since there is a flow of data through the network at each step. Ioffe et al. [59] proposed an efficient method, called Batch Normalization (BN), to alleviate this phenomenon partially. In this method, the estimators of mean and variance are calculated after each mini-batch, thereby introducing more normalization steps. Therefore, the internal covariance shift is reduced. As with this method, there is a limitation of the dependency on the initial values.

## 2.1.8. Summary

Deep CNNs have made breakthroughs in processing images, video, speech, and text. Recent CNNs are becoming deeper and deeper, requiring larger datasets and massive computing power for their training. Therefore, there has been a lot of research, either intending to reduce the complexity, without decreasing the accuracy, or to employ other strategies requiring less amount of data. Utilizing CNNs for medical image processing, computer-aided diagnosis, and decision-making systems has shown remarkable and encouraging results. CNNs require skill and experience to select suitable parameters and hyper-parameters, such as the number of filters, the number of layers, the learning rate, the parameters of the optimizers, etc. With the solid theory still lacking [26], more efforts on investigating more in-depth how and why they work is essential. This is impeding the diffusion of CNN on medical imaging and their wide application to computer-aided diagnostic systems.

## 2.2. Datasets of the study and data preprocessing

**LIDC-IDRI dataset**

The Lung Image Database Consortium and Image Database Resource Initiative (LIDC-IDRI) was introduced by the National Cancer Institute (NCI) [14]. This public dataset consists of 1010 CT scans with annotations provided by four radiologists. Each image includes one or more lung nodules. The malignancy rating of each nodule is defined by four radiologists and comes with the dataset in XML format [15].

We open the XML in Python environment, and we extract from each XML: (a) the study id, (b) the series id, (c) each nodule x, y coordinates, and (d) the malignancy rating. Using that information, we extract and crop the nodules from the CT images. We exclude nodules larger than 30mm and smaller than 3mm. To label the nodules, we consider lung nodules diagnosed by at least three radiologists, compute the median value of annotated grades for a nodule, and take a median value of less than three as benign and greater than three as malignant. We exclude all nodules with a median value equal to 3. We also exclude low-resolution nodule images and images highly doubtful as to their malignancy.

The above process results in a collection of 1050 unique lung nodules. The benign nodules are 605 and occupy 1200 images; the malignant nodules are 445 and occupy 1000 images.

LIDC IDRI dataset's labeling is weak. Nodule malignancy rating obtained from the eye of the radiologists, no matter how experienced they are, is not trustworthy. That is the reason why, in many cases, other diagnostic measurements must be performed for the characterization of the nodules (PET CT, biopsy, patient follow up). However, since, at this point of time, no public dataset exists that contains the desired labeling, we will use this dataset for further experiments.

**Dataset of our study**

Our PET/CT dataset consists of CT scans referring to 112 patient cases. 167 unique SPNs were examined from the dataset. In the original dataset, each nodule may be represented by three to fifteen images. Usually, the benign nodules occupy three to even images. The malignant nodules occupy five to fifteen.

Assuming that a random malignant nodule is represented in 10 images and the maximum size of that nodule be 3cm, the images that actually depict a nodule of size 3cm are one to three. The rest seven images depict the same nodule form a different slice position. That means that in at least four images, the nodule's size is very small and could be misclassified. This issue is not valid for the benign ones; benign nodules generally occupied no more than seven images, and the size diversity was smaller. For the above reason, we restrict the images that represent the same nodule. Each nodule is allowed to be presented by no more than seven images.

The original images were in Dicom format. The size of each image was 516kb. The info embedded into the Dicom format included information regarding the size of the image in terms of width, height, and pixels per mm: (a) width: 700mm, (b) height: 700mm, (c) pixels/mm: 0.7314. The overall pixel size of each picture was 512x512 pixels. Each patient case came with a specific medical report, including the position of the nodule(s) and their behavior (benign/malignant). All the instances were labeled. Labeling had been done by the doctors using either (a) biopsy results or, (b) FGD consumption by PET-CT scan. Instances with uncertainty as to their behavior were excluded from the dataset. Hence, we achieve accurate labeling of our dataset. The dataset includes 112 CT Scans that correspond to 112 patients. From those CT scans, 80 single benign nodules and 87 single malignant ones were extracted. Those occupy 558 and 607 images accordingly.

The nodule images were extracted in tiff format at a fixed size of 32x32 pixels. This way, we retain the nodules' characteristics. An area of 32x32 pixels is enough for a nodule of 30mm to fit in. Besides, nodules larger than 30mm are not considered solitary pulmonary nodules. The nodules were placed in the center of the picture. The reduction of the area of interest to 32x32 was supervised by a radiologist, to ensure that we don't exclude significant neighboring information. An example of benign and malignant cases, from the specific dataset, is given in Figure 3.

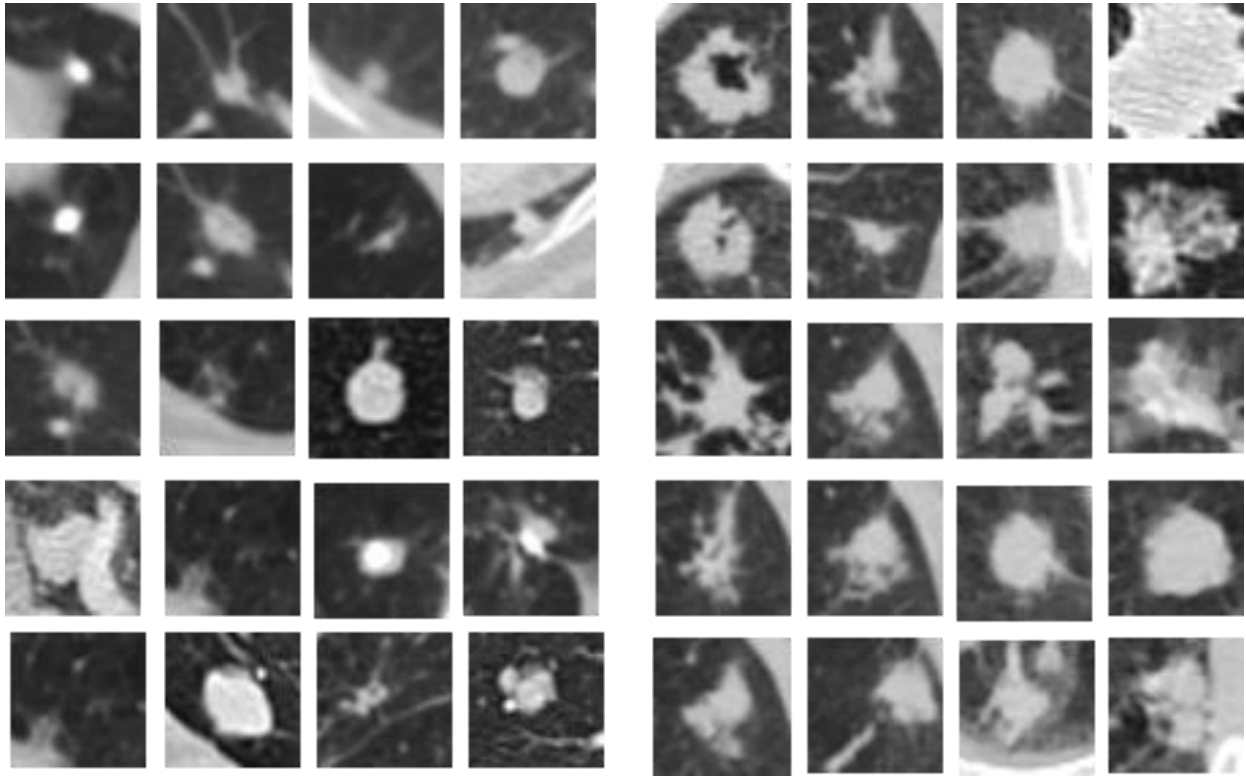

**Figure 3.** Random Solitary Pulmonary Nodules (SPN) from the PET/CT dataset

### 2.3. Data Augmentation

Data augmentation's target is to help the training by supplying the network with generated images coming from the original ones. Data augmentation should be considered and designed wisely. Feeding a learning network with unrealistic images may overcomplicate things and not provide any help.

In this study, for data augmentation, we relied on Keras' Image Data Generator function. This function is used before each training epoch to generate new images automatically. It contains several options for augmentation, including an input for a user-defined function.

We investigated the outputs of the function by saving them to a folder and letting the function generate 20.000 images. Here, we discuss the effects of several parameters of the function as follows:

- Width shift and height shift were not always beneficial to our images. Large nodules may occupy 90% of the size of the image, and therefore, shifting them results in information loss, due to their movement outside the image borders. Therefore, we reduce the range of shifting to [-3,3] pixels. We also restrict the height shift range accordingly.
- Brightness range produces too white or too black pictures, even in small ranges of brightness changes allowed.
- Rotation may also result in information loss, since pixels may be positioned outside the image borders. In the case of this experiment, we restrict the rotation scale to 30º.
- Vertical and horizontal flips do not result in information loss, as expected.

In this work, we have some forever lasting standards for our datasets that are presented as follows:

(a) The image sizes are 32x32 pixels, and the nodules may occupy every pixel. There are no pixels left for the surroundings, except for small nodules, where some neighboring information may be incorporated.

(b) The brightness and contrast of both datasets do not vary strongly.

(c) The nodules are centered.

For the above reasons, changing brightness does not produce realistic images for this work. Also, shifting images is not desirable, and rotation should be strict (maximum rotation should be around 30°). Vertical and horizontal flips should be used freely.

## 3. Experiments and results

In this section, we describe the developed models. We discuss their parameters and perform several experiments in order to tune them to the optimal ones. Specifically, the research is directed towards investigating the following CNN basic structures:

(a) Simple one-path CNN

(b) Dual-path CNN, focusing on early feature extraction

(c) Multi-path CNN, focusing on both early, and late feature extraction

For each case, several parameters are adjusted and tested, to define the optimal strategy for the specific CNN architecture. The models are trained separately on each dataset and tested on both datasets. For the evaluation of the proposed models, we use 10-fold cross-validation, and we compute the mean test accuracy over the ten folds.

The implementation of our networks is based on Keras with Tensorflow GPU as a backend [60]. For the experiments, we use a Personal Computer with processor Intel i5 9400F, NVidia RTX 2060 Super (16gb) with 8gb dedicated RAM for Cuda processes, and total RAM size of 16gb. Due to the small image sizes, and the small dataset sizes, each epoch's training time was approximately 20 seconds.

### 3.1. DeepSPN (DSPN)

DeepSPN consists of three sets of Convolution Layers, each followed by a max pooling layer. The filters are 32, 64, and 128 accordingly. The Max Pooling layers reduce the size of the input from 32x32 to 16x16, 8x8, and 4x4. A final Convolution Layer with 128 filters is then applied. The images are then flattened, and the arrays are inserted into two fully connected layers. DeepSPN is presented in Figure 4.

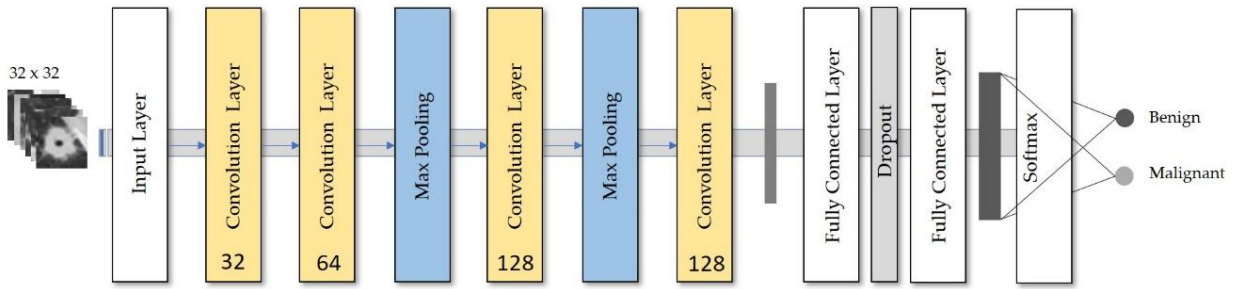

**Figure 4.** The architecture of the DeepSPN network

**Definition of the Neural Network at the top of the CNN**

For the Multi-Layer Perceptron (MLP) at the top of the network, we used two layers of 256 and 128 connections. When using one layer, the model's training accuracy reached 98%, but the test accuracy was below 82%. When using three layers of 512, 256, and 128, and with dropout layers between them, the training accuracy reached 94% and test accuracy 83%. With two layers, the training accuracy was 96%, and test accuracy 85%.

For the Dropout Layer, we used a layer that randomly disconnects 50% of the concepts. With a dropout layer of 25%, there was no significant change at the accuracy.

For the Convolution Kernel sizes, we did not stick with strictly 1x1 or 3x3 sizes. Training accuracy was below 90% when using 1x1 sizes. Sticking with 3x3 sizes resulted in rapid dimensionality reduction, and that limits the depth of the network, due to small initial image size (32x32).

Table 1 summarizes the stable parameters of the network. Further experiments were performed by adding Regularizers, substituting Dropout layers with batch normalization layers, altering the activation function, and changing the optimizers.

**Table 1.** Optimal Parameters of DeepSPN

| Stable Parameters | Description |
| --- | --- |
| Fully Connected Layers | 2 |
| Dropout | One Dropout (50%) layer |
| Kernel Sizes | Mixed 1x1 and 3x3 sizes |

### 3.1.1. Experiments on PET/CT dataset and LIDC datasets

We tune several parameters and hyperparameters of the network and compare the related accuracies. We train the model firstly on PET/CT dataset and obtain the results. Then we train it on the LIDC dataset using the optimal parameters. Table 2 presents the performance (the test accuracy) of DSPN on original and augmented datasets.

**Table 2.** Test Accuracies of DeepSPN with different hyperparameter settings

| | | Dataset for the train, and for the test | | | |
| --- | --- | --- | --- | --- | --- |
| Parameter for tuning | Other Parameters | PET/CT (%) | PET/CT Augmented (%) | LIDC (%) | LIDC augmented (%) |
| No Regularizer | Adam Optimizer, ReLU | 0.83 | 0.76 | - | - |
| Regularizer L2 at 0.01 penalty | Adam Optimizer, ReLU | 0.81 | 0.81 | - | - |
| Regularizer L2 at 0.001 | Adam Optimizer, ReLU | 0.89 | 0.90 | - | - |
| Regularizer L1 | Adam Optimizer, ReLU | 0.84 | 0.84 | - | - |
| Batch Normalization | Regularizer L2 at 0.001, ReLU, Adam | 0.82 | 0.83 | - | - |
| Global Max Pooling | Regularizer L2 at 0.001, ReLU, Adam | X | X | - | - |
| Global Average Pooling | Regularizer L2 at 0.001, ReLU, Adam | X | X | - | - |
| Optimizer SGD | Regularizer L2 at 0.001, ReLU | 0.84 | 0.86 | - | - |
| Optimizer Adam | Regularizer L2 at 0.001, ReLU | **0.89** | **0.90** | **0.92** | **0.93** |
| Optimizer Anadelta | Regularizer L2 at 0.001, ReLU | **0.89** | **0.90** | **0.92** | **0.93** |
| Activation Function ReLU | Regularizer L2 at 0.001, Adam Optimizer | **0.89** | **0.90** | **0.92** | **0.93** |
| Activation Function LeakyReLU | Regularizer L2 at 0.001, Adam Optimizer | X | X | X | X |
| Activation Function ELU | Regularizer L2 at 0.001, Adam Optimizer | X | x | x | X |

The optimal hyperparameters suggested by the results include Adam and Anadelta as optimizers, ReLU activation function, L2 kernel regularizer with a penalty set at 0.001. There were cases where the model was underfitting. Those cases are marked in the Table with an "x". Certain parameter and dataset combinations were not performed, and are marked in the Table with an "-".

We also use the trained DeepSPN on the one dataset to make predictions on the other. The results are given in Table 3.

**Table 3**. Test Accuracies of DeepSPN, when trained on the one dataset, and tested on the other.

| Training Dataset | Test Accuracy On alternative dataset (%) |
|---|---|
| PET/CT | 0.76 |
| PET/CT augmented | **0.81** |
| LIDC | 0.7 |
| LIDC augmented | 0.7 |

DeepSPN learned the training dataset adequately, achieved a level of generalization demonstrated by the high test accuracy on a sample of the dataset, but it is unable to generalize in different representations of nodules, coming from the LIDC dataset. That is a general overfitting issue.

This may be due to (a) the low number of unique nodules we use for training, (b) the diversity of image features (brightness and contrast) between different datasets, or (c) the architecture of the model is not helping it learn global characteristics. Assumptions (a) and (b) may play a role, but as we will show later, they are not the main reason behind the underperformance of DeepSPN.

The main drawback of DeepSPN is that it cannot learn the specific global features of the nodules and be able to perform at a similar level to different representations of the nodules.

The main reasons are the following:

- There is no slow introduction to a higher and deeper number of filters. In the prementioned model, the filters rapidly increase from 32 to 128, impeding the collection of both local and universal characteristics of the images.
- The images are small. We observe that after the second MaxPooling layer, the image is reduced to 8x8. This size does not allow us to apply more complex architectures and filters.
- MaxPooling, on the other hand, is vital, since it reduces the dimensions and the parameters of the system and makes it trainable form our computer. Hence, a wise combination of MaxPooling and filters must be chosen.
- Other parameters and hyperparameters could not significantly improve performance.

### 3.2. Highlights for more experimental networks

Experiments show that in order to obtain all the characteristics of an image, we must apply a vast number of filters in a way that we do not increase the parameters way too much. Building a network that requires hundreds of millions of parameters may under-fit, if not supplied with large-scale datasets.

For the next experiments, we will establish some rules, which are suggested as follows:

- **We will avoid volume collapse in the first layers**. That is, we keep the size of the image relatively close to the input dimensions. We apply a few filters in order to gather some local knowledge that may be useful.
- **We will avoid adding too many filters to the first layers for two reasons.** Firstly, to keep the amount of the trainable parameters low. (e.g. 256 filters applied on a 32x32x2 image, with valid padding and kernel size (3x3), results in an array of (30x30x256) = 230.400 parameters. Those parameters are a direct translation of local characteristics that do not necessarily collaborate with each other.
- **We will avoid sudden volume collapse due to max pooling** for the same reason mentioned above.
- **We will investigate the dropout and batch normalization techniques.** Dropout and Batch Normalization are useful to prevent overfitting, but their place in the network should be thought wisely. The high number of parameters is gathered to the FC layers, so the Dropout should be placed there.

The most straightforward way of improving the performance of deep neural networks is by increasing their size. This includes both increasing the depth of the network, and its width. Hence, it is an easy and safe way of training higher quality models, especially given the availability of a large amount of labeled training data. However, this simple solution comes with two significant drawbacks. Bigger size typically means a more significant number of parameters, which makes the extended network prone to underfitting, especially if the number of labeled examples in the training set is limited.

This can become a major bottleneck since the creation of high-quality training sets can be tricky and expensive, especially if expert human raters are necessary to distinguish between fine-grained visual categories. Another drawback of a uniformly increased network size is the dramatically increased utilization of computational resources. For example, in a deep vision network, if two convolutional layers are chained, any uniform increase in the number of their filters results in a quadratic increase of computation. If the added capacity is used inefficiently (for example, if most weights end up to be close to zero), then a lot of computation is wasted. Since, in practice, the computational budget is always finite, an efficient distribution of computing resources is preferred to an indiscriminate increase of size, even when the main objective is to increase the quality of results.

The main issue we face, when we try to follow those rules, is the following: How will we go more in-depth and apply more filters, without reducing the size of the image too much, and, simultaneously we will not produce too many parameters while keeping a slow introduction to larger filters? In the following networks, we consider a more advanced technique that contains parallel image processing tasks.

### 3.3. Dual Deep SPN (DDSPN)

DualDeepSPN is a dual-path network. The input image is independently undergoing two convolution processes. The two paths unite at one final concatenation layer, followed by one fully connected layer. The task of the first path is to gather local information and directly connect them to an FC layer. The second path's task is to take a more rapid step to denser filters and gather global characteristics.

An overview of the network is given in Figure 5. The existence of non-trainable parameters is a result of the Batch Normalization Layers distributed to the network.

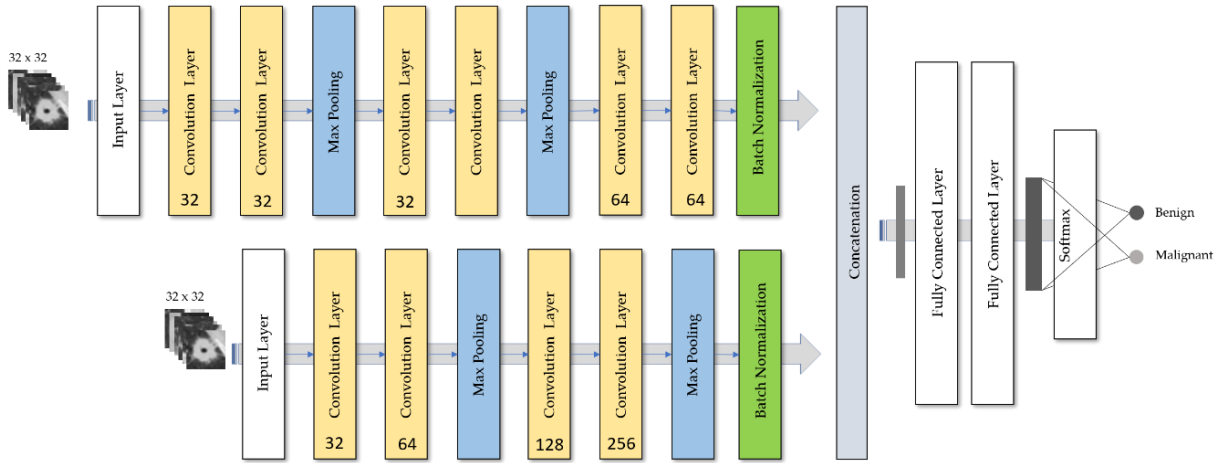

**Figure 5.** DSPN architecture

**Definition of the Neural Network at the top of the CNN**

For the Multi-Layer Perceptron (MLP) at the top of the network, we used two layers of 512 and 256 connections. When using one layer, the model's training accuracy reached 96%, but the test accuracy was 84%. When using three layers of 512, 256, and 128, with dropout layers between them, the training accuracy reached 97% and test accuracy 84%. With two layers, the training accuracy was 96%, and test accuracy 86%.

We used Batch Normalization Layer, and we removed the Dropout Layers from the MLP at the top of the network. Significant improvement was recorded in the test accuracy. For the optimized parameters, when testing on samples taken from the same dataset that was used for training, substituting Dropout Layers with Batch Normalization had no notable results. However, as we suggest in the following tables, there was an improvement to the test accuracy, when samples of the alternative datasets were given to the model for predictions. For the above reason, we stick with Batch Normalization in the specific network.

For the Convolution Kernel sizes, as we did in DeepSPN, sticking with strictly 1x1 or 3x3 sizes was avoided. As in DeepSPN, the training accuracy was below 91% when using 1x1 sizes. Sticking with 3x3 sizes resulted in rapid dimensionality reduction, and that limits the depth of the network, due to small initial image size (32x32).

Table 4 summarizes the stable parameters of the network. Further experiments were performed by adding Regularizes, substituting Dropout layers with batch normalization layers, altering the activation function, and changing the optimizers.

**Table 4.** Optimal Parameters of DualDeepSPN

| Stable Parameters | Description |
|---|---|
| Fully Connected Layers | 2 |
| Dropout | No Dropout layer |
| Kernel Sizes | Mixed 1x1 and 3x3 sizes |

### 3.3.1 Experiments on PET/CT dataset and LIDC datasets

We tune several parameters and hyperparameters of the network and compare the related accuracies. We train the model firstly on the PET/CT dataset and obtain the results by testing on the same dataset (10-fold cross-validation). Then we train it on the LIDC dataset, using the optimal parameters, and test on the LIDC dataset (10-fold cross-validation). The results are presented in Table 5.

**Table 5.** Test Accuracies of DualDeepSPN with different hyper parameter settings

| Parameter | Other Parameters | PET/CT (%) | PET/CT Augmented (%) | LIDC (%) | LIDC augmented (%) |
|---|---|---|---|---|---|
| No Regularizer | Adam Optimizer, ReLU, Batch Normalization | 0.80 | 0.75 | 0.70 | 0.71 |
| Regularizer L2 at 0.01 penalty | Adam Optimizer, ReLU, Batch Normalization | 0.80 | 0.77 | 0.71 | 0.71 |
| Regularizer L2 at 0.001 | Adam Optimizer, ReLU, Batch Normalization | 0.84 | 0.86 | 0.70 | 0.91 |
| Regularizer L1 | Adam Optimizer, ReLU, Batch Normalization | 0.83 | 0.85 | 0.69 | 0.90 |
| Dropout | Regularizer L2 at 0.001, ReLU, Adam | 0.84 | 0.85 | 0.70 | 0.90 |
| Global Max Pooling | Regularizer L2 at 0.001, ReLU, Adam | - | - | - | - |
| Global Average Pooling | Regularizer L2 at 0.001, ReLU, Adam | x | x | x | x |
| Optimizer SGD | Regularizer L2 at 0.001, ReLU, Batch Normalization | **0.84** | **0.86** | **0.70** | **0.91** |
| Optimizer Adam | Regularizer L2 at 0.001, ReLU, Batch Normalization | **0.84** | **0.86** | **0.70** | **0.91** |
| Optimizer Anadelta | Regularizer L2 at 0.001, ReLU, Batch Normalization | **0.84** | **0.86** | **0.70** | **0.91** |
| Activation Function ReLU | Regularizer L2 at 0.001, Adam Optimizer, Batch Normalization | **0.84** | **0.86** | **0.70** | **0.91** |

| Activation Function LeakyReLU | Regularizer L2 at 0.001, Adam Optimizer, Batch Normalization | x | x | x | x |
| Activation Function ELU | Regularizer L2 at 0.001, Adam Optimizer, Batch Normalization | x | x | x | X |

The optimal hyperparameters suggested by the results, include Adam and Anadelta as optimizers, ReLU activation function, L2 kernel regularizer with a penalty set at 0.001. There were cases where the model was underfitting. Those cases are marked in the Table with an "x". Some parameters were not tested in the above experiments, and are marked with a dash ("- ") in the above table.

We also use the trained DualDeepSPN on the one dataset to make predictions on the other. The results are given in Table 6.

**Table 6**. Test Accuracies of DeepSPN.

| Training Dataset | Test Accuracy On alternative dataset |
|---|---|
| PET/CT | 0.77 |
| PET/CT augmented | **0.80** |
| LIDC | 0.60 |
| LIDC augmented | **0.91** |

DDSPN is achieving acceptable results regarding its ability to generalize. We observe that when the model was trained on either augmented dataset, it achieves better accuracies than DeepSPN. That is, when trained on the augmented LIDC dataset, the test accuracy on PET/CT is 80%. When trained on the augmented PET/CT dataset, the accuracy on the LIDC dataset is 91%, which is the same accuracy the model obtained when also trained on the augmented LIDC dataset and tested on the LIDC dataset, with 10-fold cross-validation.

It is notable that the DDSPN could not achieve test accuracy above 90% on the PET/CT dataset in any experiment, except for the case where the CNN was trained on the augmented LIDC dataset. The above experiment suggests that the augmented LIDC dataset could be used as a trainer of the networks in future experiments.

3.4. ThreeDeepSPN (TDSPN)

This network is similar to DualDeepSPN but has an extra Convolution process path that is applied directly to the initial input image. The overall architecture is given in the image below. Its architecture is presented in Figure 6.

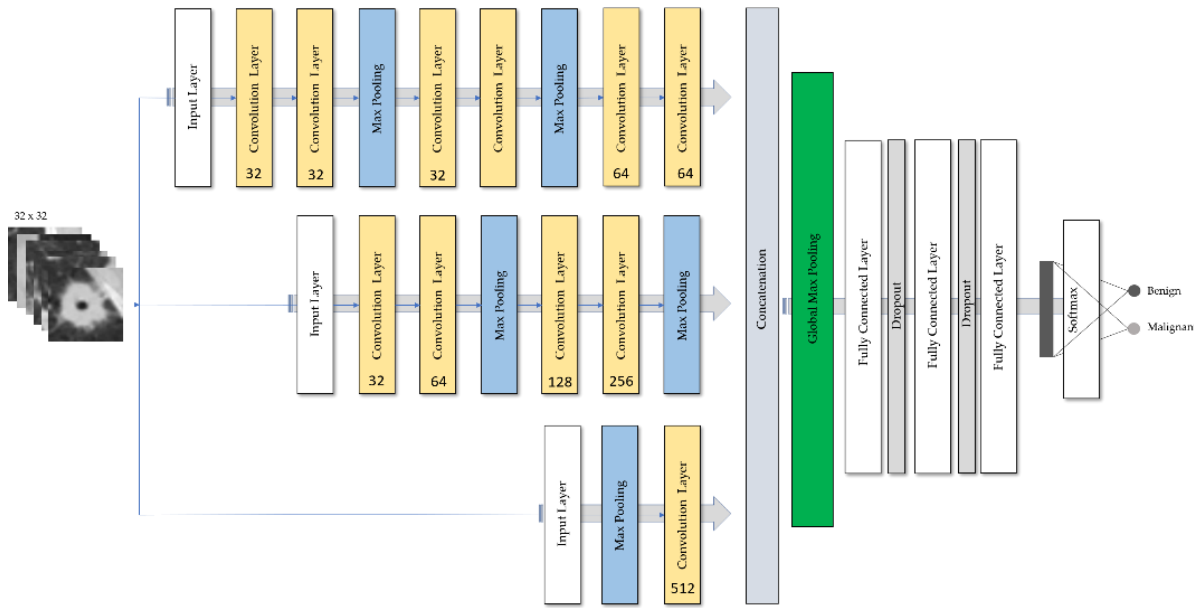

**Figure 6.** ThreeDeepLIDC architecture

We use valid padding in order to inspect the dimensionality reduction taking place upon every convolution layer. The first path collects local characteristics of the image, applying a slow introduction to larger numbers of filters. The convolution kernels are strictly 3x3, in order to have a slow dimensionality reduction, before the new characteristics are learned. In order to apply more filters, we add two MaxPooling Layers between the Convolutional Layers.

The second path is applying a more rapid introduction to 256 filters, and its architecture is similar to DeepSPN. The 256-filter Convolutional Layer is applied on a 14x14 image. After the Convolution, a MaxPooling layer is added, to reduce the dimensions to 4x4.

The third path is a shallow-type process. A Max Pooling layer of 5x5 kernel is applied to the input image, reducing its size to 6x6. Then, a 512-filter Convolution Layer is applied.

After the Layer Concatenation (i.e., the merge of the three paths), a GlobalMaxPooling layer is applied, to reduce overfitting. GlobalMaxPooling is introducing heavyweight penalties, and it replaces Dropout and Batch Normalization. Notice that the image is not flattened before the Fully Connected Layers. That is due to the existence of GlobalMaxPooling.

Table 7 summarizes the stable parameters of the network. Further experiments were performed by adding Regularizers, substituting Dropout layers with batch normalization layers, altering the activation function, and changing the optimizers.

**Table 7.** Optimal Parameters of ThreeDeepLIDC

| Stable Parameters | Description |
| --- | --- |
| Fully Connected Layers | No |
| Dropout | No Dropout layer |
| Kernel Sizes | 1x1, 3x3, and 5x5 |

### 3.4.1 Experiments on PET/CT dataset and LIDC datasets

We tune several parameters and hyperparameters of the network and compare the related accuracies. We train the model firstly on PET/CT dataset and obtain the results by testing on the same dataset (10-fold cross-validation). Then we train it on the LIDC dataset using the optimal parameters and test on LIDC (10-fold cross-validation). The results are given in Table 8.

**Table 8.** Test Accuracies of DualDeepSPN with different hyper parameter settings

| Parameter for tuning / | Other Parameters | PET/CT | PET/CT Augmented | LIDC | LIDC augmented |
|---|---|---|---|---|---|
| No Regularizer | Adam Optimizer, ReLU, Batch Normalization | - | - | - | - |
| Regularizer L2 at 0.01 penalty | Adam Optimizer, ReLU, Batch Normalization | - | - | - | - |
| Regularizer L2 at 0.001 | Adam Optimizer, ReLU, Batch Normalization | 0.80 | 0.79 | 0.81 | 0.81 |
| Regularizer L1 | Adam Optimizer, ReLU, Batch Normalization | 0.79 | 0.80 | 0.81 | 0.80 |
| Global Average Pooling | Regularizer L2 at 0.001, ReLU | x | x | X | x |
| Global Max Pooling | Regularizer L2 at 0.001, ReLU | 0.79 | 0.79 | 0.84 | 0.84 |
| Global Max Pooling | Regularizer L2 at 0.001, ELU | **0.82** | **0.81** | **0.89** | **0.87** |
| Optimizer SGD | Regularizer L2 at 0.001, ReLU, GlobalMaxPooling | 0.81 | 0.81 | 0.89 | 0.84 |
| Optimizer Adam | Regularizer L2 at 0.001, ReLU, GlobalMaxPooling | **0.82** | **0.81** | **0.89** | **0.87** |
| Optimizer Anadelta | Regularizer L2 at 0.001, ReLU, GlobalMaxPooling | **0.82** | **0.81** | **0.89** | **0.87** |
| Activation Function ReLU | Regularizer L2 at 0.001, Adam Optimizer, Batch Normalization | 0.80 | 0.77 | 0.81 | 0.85 |
| Activation Function ReLU | Regularizer L2 at 0.001, Adam Optimizer, GlobalMaxPooling | 0.81 | X | 0.74 | x |
| Activation Function Leaky ReLU | Regularizer L2 at 0.001, Adam Optimizer, Batch Normalization | **0.80** | x | **0.85** | x |

The optimal hyperparameters suggested by the results include Adam and Anadelta as optimizers, ELU activation function, L2 kernel regularizer with a penalty set at 0.001. There were cases where the model was underfitting. Those cases are marked in the Table with an "x". Some parameters were not tested in the above experiments, and are marked with a dash ("- ") in the above table.

We observed the benefits of ELU activation function and GlobalMaxPooling. We experimented more by trying the combination of ELU and GlobalMaxPooling to the previous models (DeepSPN and DualDeepSPN). Some observations include the following:

- When the previous models were tested again, but with the ELU activation function, they did not improve their accuracy (in fact, we got a 4% decrease as the previous Tables suggest). When tested with GlobalMaxPooling and ELU, the training process was stopped, due to underfitting.
- When Global Average Pooling was used instead of GlobalMaxPooling, the model could not reach training accuracy above 80%. However, its test accuracy was above 80% on both datasets. That underlines that Global Average Pooling is severely reducing overfitting, but the model is on the verge of underfitting.
- The three paths of the model work together, but do not give acceptable results individually. We observed a decrease in test accuracy when every path was trained and tested alienated from the other paths.

We also use the trained ThreeDeepLIDC for transfer learning. As we did for DDSPN, we do not fine-tune the pretrained network. We freeze every parameter. The results are given in Table 9.

**Table 9**. Test Accuracies of DeepSPN, used for transfer learning

| Training Dataset | Test Accuracy On alternative dataset |
|---|---|
| PET/CT | 0.80 |
| PET/CT augmented | 0.73 |
| LIDC | 0.69 |
| LIDC augmented | 0.72 |

## 4. Discussion

Overfitting was the major issue we had to tackle during the experiments. Overfitting prevention methods such as Dropout layers, Batch Normalization layers, Data augmentation, and Regularizers proved useful as the models improved their test accuracy. There was diversity as to which method benefited which model. DSPN was not benefited from the Batch Normalization layer, and Dropout layers were preferred. DDSPN, on the other side, worked better with Batch Normalization. Both DSPN and DDSPN were underfitting when Global Max Pooling was applied, though TDLIDC was benefited by it.

Data augmentation also had diverse effects on the CNNs. What is more, we observed different behavior of the same CNNs when augmentation was applied to both the datasets. More specifically, DSPN only increased its accuracy by 1% when trained with augmented data. DDSPN was benefitted from augmenting both datasets, having a maximum increase in accuracy by 21%. THLIDC's accuracy was decreased on the augmented LIDC dataset by 1%, whereas the accuracy was increased on augmented PET/CT.

We observed that the utilization of ELU activation function benefited only THLIDC, by improving its accuracy by 3%-5% on either dataset. DSPN and DDSPN were underfitting when ELU was substituted with ReLU.

Though overfitting was prevented, none of the models learned what a nodule is. This is demonstrated by the model's accuracy when they were employed for transfer learning. Trained on either dataset, most of the CNNs could not correctly classify more than 80% of the alternative dataset of the same nature. DDSPN reached 91% accuracy on PET/CT when trained on LIDC; however, when switching the training and test sets and retrained, the accuracy decreased. That suggests that (a) compared to the other model, DDSPN could generalize more, even though it was not the best architecture for the initial classification task, (b) the LIDC dataset was an excellent trainer both for itself and for the PET/CT dataset.

## 5. Conclusions

In this work, we considered the problem of Solitary Pulmonary Nodule characterization, using Deep Learning. More specifically, we employed different Convolutional Network Architectures and performed experiments by tuning the CNN parameters and experimenting with different hyperparameters. As shown in section 2, there are numerous improvements proposed recently, which increase CNNs capabilities with respect to the task they are employed to complete. The analytical inspection of every proposal is impossible and exceeds the purpose of this work.

We designed three different Network architectures and performed experiments using the most common and famous parameter proposals regarding the activation function, the Multilayer Perceptron at the top of the networks, the regularizers, Pooling layers, and methods for overfitting prevention. We also used data augmentation during the experiments in order to specify the affection of augmented data in different network architectures.

For the preparation of our data, specific data preprocessing steps were performed. Image and label extraction from the LIDC dataset and PET/CT dataset were followed by the area of interest reduction and crop.

Concluding, we showed that hyper-parameter tuning can benefit a CNN, but cannot transform it from an ineffective model to an effective one. More wisely should the initial architectures be designed, rather than the definition of the other parameters. Finally, we obtain an acceptable (above 90%) SPN classification rate for both datasets. Further research should look towards enhancing the learning capacity of the CNNs in order to improve the robustness of the networks.

## References

1. Torre, L.A. et al.: Lung Cancer Statistics. In: Ahmad, A. and Gadgeel, S. (eds.) Lung Cancer and Personalized Medicine. pp. 1–19 Springer International Publishing, Cham (2016). https://doi.org/10.1007/978-3-319-24223-1_1.

2. Siegel, R. et al.: Cancer statistics, 2014: Cancer Statistics, 2014. CA A Cancer Journal for Clinicians. 64, 1, 9–29 (2014). https://doi.org/10.3322/caac.21208.

3. Zia ur Rehman, M. et al.: An appraisal of nodules detection techniques for lung cancer in CT images. Biomedical Signal Processing and Control. 41, 140–151 (2018). https://doi.org/10.1016/j.bspc.2017.11.017.

4. Renfrew, D.L. et al.: Error in radiology: classification and lessons in 182 cases presented at a problem case conference. Radiology. 183, 1, 145–150 (1992). https://doi.org/10.1148/radiology.183.1.1549661.

5. Zhu, W. et al.: DeepLung: Deep 3D Dual-Path Nets for Automated Pulmonary Nodule Detection and Classification. arXiv:1801.09555 [cs]. (2018).

6. Doi, K.: Diagnostic imaging over the last 50 years: research and development in medical imaging science and technology. Physics in Medicine & Biology. 51, 13, R5 (2006).

7. Deng, L.: Deep Learning: Methods and Applications. FNT in Signal Processing. 7, 3–4, 197–387 (2014). https://doi.org/10.1561/2000000039.

8.Greenspan, H. et al.: Guest editorial deep learning in medical imaging: Overview and future promise of an exciting new technique. IEEE Transactions on Medical Imaging. 35, 5, 1153–1159 (2016).

9. Szegedy, C. et al.: Going deeper with convolutions. In: Proceedings of the IEEE conference on computer vision and pattern recognition. pp. 1–9 (2015).

10. Krizhevsky, A. et al.: ImageNet classification with deep convolutional neural networks. Commun. ACM. 60, 6, 84–90 (2017). https://doi.org/10.1145/3065386.

11. Simonyan, K., Zisserman, A.: Very Deep Convolutional Networks for Large-Scale Image Recognition. arXiv:1409.1556 [cs]. (2015).

12. Howard, A.G. et al.: MobileNets: Efficient Convolutional Neural Networks for Mobile Vision Applications. arXiv:1704.04861 [cs]. (2017).

13. Szegedy, C. et al.: Inception-v4, inception-resnet and the impact of residual connections on learning. In: Thirty-First AAAI Conference on Artificial Intelligence. (2017).

13. Han, F. et al.: Texture Feature Analysis for Computer-Aided Diagnosis on Pulmonary Nodules. J Digit Imaging. 28, 1, 99–115 (2015). https://doi.org/10.1007/s10278-014-9718-8.

14. Armato, S.G. et al.: The Lung Image Database Consortium (LIDC) and Image Database Resource Initiative (IDRI): A Completed Reference Database of Lung Nodules on CT Scans: The LIDC/IDRI thoracic CT database of lung nodules. Med. Phys. 38, 2, 915–931 (2011). https://doi.org/10.1118/1.3528204.

15. Armato, S.G. et al.: LUNGx Challenge for computerized lung nodule classification. J. Med. Imag. 3, 4, 044506 (2016). https://doi.org/10.1117/1.JMI.3.4.044506.

16. Wang, X. et al.: An Appraisal of Lung Nodules Automatic Classification Algorithms for CT Images. Sensors. 19, 1, 194 (2019). https://doi.org/10.3390/s19010194.

17. LeCun, Y. et al.: Handwritten Digit Recognition with a Back-Propagation Network. In: Advances in Neural Information Processing Systems 2, [NIPS Conference, Denver, Colorado, USA, November 27-30, 1989]. pp. 396–404 (1989).

18. LeCun, Y. et al.: Deep learning. Nature. 521, 7553, 436–444 (2015). https://doi.org/10.1038/nature14539.

19. LeCun, Y. et al.: Convolutional networks and applications in vision. In: Proceedings of 2010 IEEE International Symposium on Circuits and Systems. pp. 253–256 IEEE, Paris, France (2010). https://doi.org/10.1109/ISCAS.2010.5537907.

20. Liu, M. et al.: Towards Better Analysis of Deep Convolutional Neural Networks. arXiv:1604.07043 [cs]. (2016).

21. Kalchbrenner, N. et al.: Neural machine translation in linear time. arXiv preprint arXiv:1610.10099. (2016).

22. Ngiam, J. et al.: Tiled convolutional neural networks. In: Advances in neural information processing systems. pp. 1279–1287 (2010).

23. Zeiler, M.D., Fergus, R.: Visualizing and understanding convolutional networks. In: European conference on computer vision. pp. 818–833 Springer (2014).

24. Lin, M. et al.: Network in network. arXiv preprint arXiv:1312.4400. (2013).

25. Nair, V., Hinton, G.E.: Rectified linear units improve restricted boltzmann machines. In: Proceedings of the 27th international conference on machine learning (ICML-10). pp. 807–814 (2010).

26. Russakovsky, O. et al.: Imagenet large scale visual recognition challenge. International journal of computer vision. 115, 3, 211–252 (2015).

27. Clevert, D.-A. et al.: Fast and accurate deep network learning by exponential linear units (elus). arXiv preprint arXiv:1511.07289. (2015).

28. Maas, A.L. et al.: Rectifier nonlinearities improve neural network acoustic models. In: Proc. icml. p. 3 (2013).

29. He, K. et al.: Delving deep into rectifiers: Surpassing human-level performance on imagenet classification. In: Proceedings of the IEEE international conference on computer vision. pp. 1026–1034 (2015).

30. Xu, B. et al.: Empirical evaluation of rectified activations in convolutional network. arXiv preprint arXiv:1505.00853. (2015).

31. Goodfellow, I.J. et al.: Maxout networks. arXiv preprint arXiv:1302.4389. (2013).

32. Springenberg, J.T., Riedmiller, M.: Improving deep neural networks with probabilistic maxout units. arXiv preprint arXiv:1312.6116. (2013).

33. Hyvärinen, A., Köster, U.: Complex cell pooling and the statistics of natural images. Network: Computation in Neural Systems. 18, 2, 81–100 (2007).

34. Yu, D. et al.: Mixed pooling for convolutional neural networks. In: International Conference on Rough Sets and Knowledge Technology. pp. 364–375 Springer (2014).

35. Zeiler, M.D., Fergus, R.: Stochastic pooling for regularization of deep convolutional neural networks. arXiv preprint arXiv:1301.3557. (2013).

36. Rippel, O. et al.: Spectral representations for convolutional neural networks. In: Advances in neural information processing systems. pp. 2449–2457 (2015).

37. He, K. et al.: Spatial pyramid pooling in deep convolutional networks for visual recognition. IEEE transactions on pattern analysis and machine intelligence. 37, 9, 1904–1916 (2015).

38. Gong, Y. et al.: Multi-scale orderless pooling of deep convolutional activation features. In: European conference on computer vision. pp. 392–407 Springer (2014).

39. Bottou, L.: Stochastic Gradient Learning in Neural Networks. In: Proceedings of Neuro-Nîmes 91. EC2, Nimes, France (1991).

40. Zhang, T.: Solving large scale linear prediction problems using stochastic gradient descent algorithms. In: Proceedings of the twenty-first international conference on Machine learning. p. 116 ACM (2004).

41. Liu, W. et al.: Large-margin softmax loss for convolutional neural networks. In: ICML. p. 7 (2016).

42. Bromley, J. et al.: Signature verification using a" siamese" time delay neural network. In: Advances in neural information processing systems. pp. 737–744 (1994).

43. Schroff, F. et al.: Facenet: A unified embedding for face recognition and clustering. In: Proceedings of the IEEE conference on computer vision and pattern recognition. pp. 815–823 (2015).

44. Vincent, P. et al.: Extracting and composing robust features with denoising autoencoders. In: Proceedings of the 25th international conference on Machine learning. pp. 1096–1103 ACM (2008).

45. Hinton, G.E. et al.: Improving neural networks by preventing co-adaptation of feature detectors. arXiv preprint arXiv:1207.0580. (2012).

46. Tikhonov, A.N.: On the stability of inverse problems. In: Dokl. Akad. Nauk SSSR. pp. 195–198 (1943).

47. Wang, S., Manning, C.: Fast dropout training. In: international conference on machine learning. pp. 118–126 (2013).

48. Ba, J., Frey, B.: Adaptive dropout for training deep neural networks. In: Advances in Neural Information Processing Systems. pp. 3084–3092 (2013).

49. Tompson, J. et al.: Efficient object localization using convolutional networks. In: Proceedings of the IEEE Conference on Computer Vision and Pattern Recognition. pp. 648–656 (2015).

50. Yang, H., Patras, I.: Mirror, mirror on the wall, tell me, is the error small? In: Proceedings of the IEEE Conference on Computer Vision and Pattern Recognition. pp. 4685–4693 (2015).

51. Xie, S., Tu, Z.: Holistically-nested edge detection. In: Proceedings of the IEEE international conference on computer vision. pp. 1395–1403 (2015).

52. Salamon, J., Bello, J.P.: Deep convolutional neural networks and data augmentation for environmental sound classification. IEEE Signal Processing Letters. 24, 3, 279–283 (2017)

53. Eigen, D., Fergus, R.: Predicting depth, surface normals and semantic labels with a common multi-scale convolutional architecture. In: Proceedings of the IEEE international conference on computer vision. pp. 2650–2658 (2015).

54. Goodfellow, I. et al.: Generative adversarial nets. In: Advances in neural information processing systems. pp. 2672–2680 (2014).

55. Chawla, N.V. et al.: SMOTE: synthetic minority over-sampling technique. Journal of artificial intelligence research. 16, 321–357 (2002).

56. Mishkin, D., Matas, J.: All you need is a good init. arXiv preprint arXiv:1511.06422. (2015).

57. Glorot, X., Bengio, Y.: Understanding the difficulty of training deep feedforward neural networks. In: Proceedings of the thirteenth international conference on artificial intelligence and statistics. pp. 249–256 (2010).

58. Saxe, A.M. et al.: Dynamics of learning in deep linear neural networks. In: NIPS Workshop on Deep Learning. (2013).

59. Ioffe, S., Szegedy, C.: Batch normalization: Accelerating deep network training by reducing internal covariate shift. arXiv preprint arXiv:1502.03167. (2015).

60. Chollet, F.: keras. GitHub (2015).

**Acknowledgements**

We thank the Laboratory of Nuclear Medicine of the University Hospital of Patras for approving access to the image database and providing this study with anonymous CT scans.